\documentclass[reprint,superscriptaddress,amsmath,amssymb,aps,prl]{revtex4-1}
\usepackage{graphicx}
\usepackage{dcolumn}
\usepackage{bm}
\usepackage[colorlinks,allcolors=blue]{hyperref}

\begin{document}

\preprint{}

\title{Record high $T_\mathrm{c}$ and robust superconductivity in transition metal $\delta$-Ti phase at megabar pressure}

\author{Xuqiang Liu}
\thanks{The equal contribution to this work}
\affiliation{Key Laboratory for Anisotropy and Texture of Materials, Northeastern University, Shenyang 110819, China}
\affiliation{Center for High Pressure Science and Technology Advanced Research (HPSTAR), Shanghai 201203, China}

\author{Peng Jiang}
\thanks{The equal contribution to this work}
\affiliation{School of Physics and Electronic Engineering, Jiangsu Normal
University, Xuzhou 221116, China}

\author{Yiming Wang}
\affiliation{Center for High Pressure Science and Technology Advanced Research
(HPSTAR), Shanghai 201203, China}

\author{Mingtao Li}
\affiliation{Center for High Pressure Science and Technology Advanced Research
(HPSTAR), Shanghai 201203, China}

\author{Nana Li}
\affiliation{Center for High Pressure Science and Technology Advanced Research
(HPSTAR), Shanghai 201203, China}

\author{Qian Zhang}
\affiliation{Center for High Pressure Science and Technology Advanced Research (HPSTAR),
Shanghai 201203, China}
\affiliation{School of Materials and Chemical Engineering, Zhongyuan University of Technology,
Zhengzhou 451191, China}

\author{Yandong Wang}
\affiliation{Key Laboratory for Anisotropy and Texture of Materials, Northeastern University, Shenyang 110819, China}

\author{Yan-ling Li}
\email{ylli@jsnu.edu.cn}
\affiliation{School of Physics and Electronic Engineering, Jiangsu Normal
University, Xuzhou 221116, China}

\author{Wenge Yang}
\email{yangwg@hpstar.ac.cn}
\affiliation{Center for High Pressure Science and Technology Advanced Research (HPSTAR),
Shanghai 201203, China}

\date{\today}

\begin{abstract}
We report a record high superconducting transition temperature ($T_\mathrm{c}$) up to 23.6~K under high pressure in 
the elemental metal Ti, one of the top ten most abundant elements in Earth's crust. 
The $T_\mathrm{c}$ increases monotonically from 2.3 K at 40.3~GPa to 23.6~K at 144.9~GPa, which surpasses all known records from 
elemental metals reported so far. With further compression, a robust $T_\mathrm{c}$ of $\sim$23 K is observed between 144.9 and 183~GPa 
    in the $\delta$-Ti phase. The pressure-dependent $T_\mathrm{c}$ can be well described by the conventional electron-phonon coupling (EPC) mechanism. 
Density Functional Theory calculations show the Fermi nesting and the phonon softening of optical branches at the $\gamma$-Ti to $\delta$-Ti 
phase transition pressure enhance EPC, which results in the record high $T_\mathrm{c}$. We attribute the robust superconductivity in 
$\delta$-Ti to the apparent robustness of its strong EPC against lattice compression. 
These results provide new insight into exploring new high-$T_\mathrm{c}$ elemental metals and Ti-based superconducting alloys.
\end{abstract}

\maketitle
Discovering materials with a high $T_\mathrm{c}$ is an active interest in condensed matter physics 
\cite{PhysRevLett.58.908,Drozdov2015,PhysRevLett.122.027001,Snider2020,Zhang2017,doi:10.1063/5.0040607}. 
Simple superconducting elements are the original and most suitable platform to testify 
the Bardeen-Cooper-Schrieffer (BCS) theory
\cite{PhysRev.167.331,PhysRevB.12.905}.
To date, over fifty elements at ambient and high pressure have been discovered
to host superconductivity \cite{SHIMIZU201830}, and more attention is especially paid to the
transition metals (TMs). At ambient conditions, most TMs with partially filled
$d$-orbitals are superconductors \cite{HAMLIN201559}. By applying pressure, a remarkable
increase of $T_\mathrm{c}$ has been found in some TMs such as scandium (Sc)
\cite{PhysRevB.78.064519,PhysRevLett.94.195503,PhysRevB.72.132103,PhysRevB.73.134102},
yttrium (Y) \cite{HAMLIN200782,PhysRevB.102.094104,PhysRevLett.109.157004,PhysRevB.99.085137}, 
and vanadium (V) \cite{PhysRevB.61.R3823,PhysRevLett.98.085502,Suzuki_2002}. 
Beyond TMs, calcium (Ca) is believed so far to have the highest $T_\mathrm{c}$ near 21~K 
(accompanied by a superconductivity fluctuation at 29~K) among all elemental
metals at $\sim$216~GPa, where Ca-VI (Pnma) transforms to Ca-VII (host-guest structure) 
\cite{HAMLIN201559,PhysRevB.83.220512,PhysRevLett.110.235501,yabuuchi2006superconductivity}. 
The underlying mechanism of pressure-enhanced $T_\mathrm{c}$ in Ca \cite{GAO2008181,Aftabuzzaman_2011}, 
Sc \cite{PhysRevB.76.134512,Bose_2008}, Y\cite{PhysRevLett.109.157004,PhysRevB.75.024512}, 
and V \cite{PhysRevB.67.094509} has been explained by electron-phonon coupling (EPC) or
spin fluctuation, which is closely associated with the common $s$-$d$ electron
transfer \cite{PhysRevB.31.1909,mcmahan1986pressure,Pettifor_1977,PhysRevB.62.12743}. 
Their maximum $T_\mathrm{c}$ ($T^\mathrm{max}_\mathrm{c}$) probably correlates with the completion
degree of the $s \rightarrow d$ transfer. For Ca, $T^\mathrm{max}_\mathrm{c}$ appears in a complex host-guest
structure \cite{PhysRevLett.110.235501}, similar to the Ba-VI structure with the near completion of the
$s \rightarrow d$ transfer \cite{loa2012extraordinarily}. 
A study of the $T_\mathrm{c}$-dependent number of $d$-electrons in the conduction band ($N_d$)
for Sc and Y develops a phenomenological model where $T_\mathrm{c}$ approaches a saturated
value once the $s \rightarrow d$ transfer completes as the $N_d$ $\rightarrow$ 3 rule \cite{PhysRevB.78.064519}. 
It is also theoretically reported that $T^\mathrm{max}_\mathrm{c}$ appears at
$N_d$ $\sim$ 4 in V under pressure
of 139.3~GPa, and $T_\mathrm{c}$ then decreases as $N_d$ approaches 5 with the half-filled nature of its $d$-orbital 
\cite{PhysRevB.67.094509}. Hence, one intuitively expects that the pressure-induced $s \rightarrow d$ transfer in group IVB TMs with the electronic
configuration $nd^2(n+1)s^2$ may reach a considerably high $T_\mathrm{c}$.

As one of the group IVB TMs adjacent to the high-$T_\mathrm{c}$ Ca, Sc, V and Y, pressurized titanium (Ti) undergoes 
a structural transition sequence: $\alpha$ $(P63/mmc)$ $\rightarrow$
$\omega$ $(P6/mmm)$ $\rightarrow$ $\gamma$ $(Cmcm)$ $\rightarrow$ $\delta$ $(Cmcm)$
$\rightarrow$ $\beta$ $(Im3m)$ \cite{PhysRevLett.87.275503,PhysRevLett.86.3068,PhysRevB.67.132102,
PhysRevB.69.184102,HAO20101473,PhysRevB.91.134108,doi:10.1063/5.0014766},
where the $\gamma$ and $\delta$ phases do not occur in pressurized zirconium (Zr) 
\cite{PhysRevLett.64.204,jamieson1963crystal,akahama1991high} and hafnium
(Hf) \cite{PhysRevB.42.6736,Gyanchandani_1990}. Unlike the $\alpha$ and
$\omega$ phases, the $\beta$ phase in Zr and Hf has a
negative slope of $dT_\mathrm{c}$/$dP$ \cite{akahama1991high,doi:10.1143/JPSJ.59.3843}. 
The $T^\mathrm{max}_\mathrm{c}$ of $\beta$-Zr appears at 33~GPa with $N_d$ = 3.5
\cite{akahama1991high}. This leads us to infer the occurrence of the $\beta$ phase in group IVB TMs
signals the completion of the $s \rightarrow d$ transfer, simultaneously
triggering a $T^\mathrm{max}_\mathrm{c}$. For Ti, interestingly, the $\gamma$ and
$\delta$ phases sequentially appear in a large pressure interval (over
100~GPa) before transforming into $\beta$ phase \cite{PhysRevLett.87.275503,PhysRevLett.86.3068}.
Thus, we expect that the broad interval is a promising fertile ground for
obtaining high-$T_\mathrm{c}$ superconductivity in Ti. Indeed the
$T_\mathrm{c}$ of $\omega$-Ti was reported to increase from 2.3~K at 40.9~GPa
to 3.4~K at 56.0~GPa \cite{BASHKIN200712}. 
However, such a small positive $dT_\mathrm{c}$/$dP$ has not triggered further transport
measurements at higher pressures. Up to now, the $T_\mathrm{c}$ in the
$\gamma$-Ti and $\delta$-Ti phases remains absent. This motivated us to extend the transport measurements
of Ti beyond megabar pressure.

This letter presents a comprehensive study of the superconducting behavior up
to the $\delta$-Ti phase near 2 Mbar. Interestingly, our results show that the
$T^\mathrm{max}_\mathrm{c}$ reaches 23.6~K at a pressure of about 145~GPa, renewing
the record value in TMs. After that, the $T_\mathrm{c}$ becomes nearly saturated in the
pressure range of 145-183~GPa, manifesting robust superconductivity.
Furthermore, theoretical calculations identify that the conventional EPC
mechanism can capture the evolution of the $T_\mathrm{c}$ in pressurized Ti well.

High-pressure electrical transport measurements with a four-probe
configuration were performed with a physical property measurements system
(PPMS, DynaCool, Quantum Design Inc.). A BeCu alloy diamond anvil cell with
100~$\mu$m diameter culets was employed to generate pressure beyond Megabar. The
pressure was determined by using the diamond Raman method
\cite{doi:10.1063/1.2335683}. A rhenium gasket and powder of cubic boron-nitride mixed with epoxy were used to create
the sample chamber and electric insulation. A thin titanium foil (Alfa Aesar,
99.99\%) was used as a sample placed on top of four thin Pt probes. Details of
our theoretical methods are described in Supplemental Material (SM) \cite{supp}, including references
\cite{oganov2011evolutionary,lyakhov2013new,kresse1996efficient,kresse1999ultrasoft,blochl1994projector,
giannozzi2009quantum,baroni2001phonons,bardeen1957theory,hamann2013optimized}.

\begin{figure}[thp]
        \includegraphics[width=0.40 \textwidth]{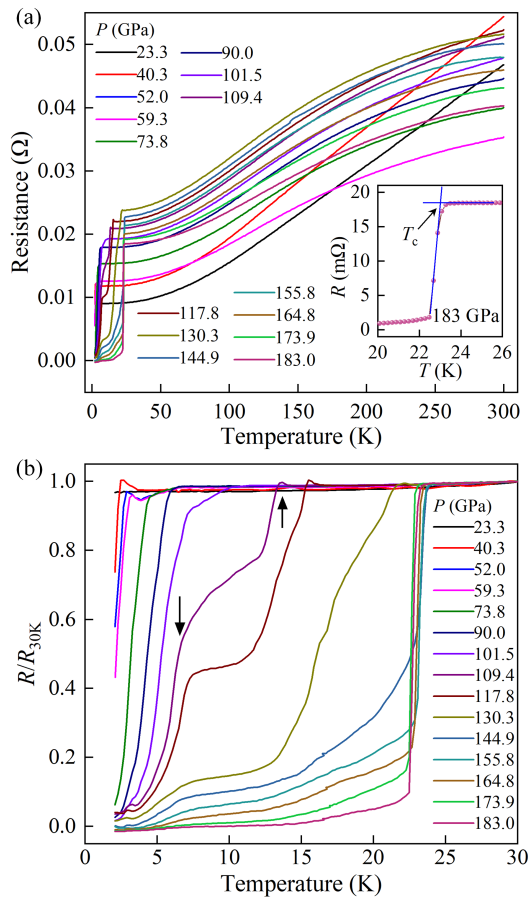}
            \caption{(a) Temperature-dependent resistance
            $R(T)$ of Ti from 2 to 300~K at pressure up to 183~GPa. Inset shows
            the determination of the $T_\mathrm{c}$ at 183~GPa. (b) Zoom-in view of the
            normalized resistance in the low temperatures region from
            Fig.~\ref{fig1}(a), clearly displays the pressure effect on the
            $T_\mathrm{c}$. The arrows indicate the two superconducting phases coexisting 
            at 109.4~GPa, associated with the phase transition of $\omega$
            $\rightarrow$ $\gamma$.}\label{fig1}
\end{figure}

The temperature-dependent resistance $R(T)$ measurements up to 183~GPa are
plotted in Fig.~\ref{fig1}(a). All $R(T)$ data show metallic behavior in the normal
state. The bottom-right displays a representative definition of the
$T_\mathrm{c}$ at 183~GPa, in which the intersection signals the superconducting transition at 
$T_\mathrm{c}$ = 23~K. The zoom-in view of the $R(T)$ in the low-temperature
region (Fig.~\ref{fig1}(b)) shows a sharp drop occurring at 2.3~K and 40.3~GPa. This result is in line
with the previous report \cite{BASHKIN200712}. The $T_\mathrm{c}$ shifts to a higher temperature with
increasing pressure. Two noticeable drops in $R(T)$ are observed at 109.4~GPa,
indicating the coexistence of two superconducting phases. The magnetic field
suppression of the superconducting transition in Fig.~S1 of the SM \cite{supp} shows two distinct slopes 
$dH_{c2}$/$dT_\mathrm{c}$, which further supports the individual phases. At
130.3~GPa, the drop at a relatively lower temperature was gradually suppressed with increasing pressure. 
Based on previous studies
\cite{PhysRevLett.87.275503,PhysRevLett.86.3068,PhysRevB.67.132102,
HAO20101473,PhysRevB.91.134108,PhysRevB.75.014109,HAO2008105}, the phase transition regions in
Ti of $\omega$ $\rightarrow$ $\gamma$ and $\gamma$ $\rightarrow$ $\delta$ were determined experimentally and 
theoretically to be 90-128~GPa and 106-140~GPa, respectively. Therefore, most of the sample had
already transformed to $\delta$-Ti, and the robust superconductivity beyond 140~GPa
is within the $\delta$-Ti phase.

\begin{figure}[thp]
        \includegraphics[width=0.40 \textwidth]{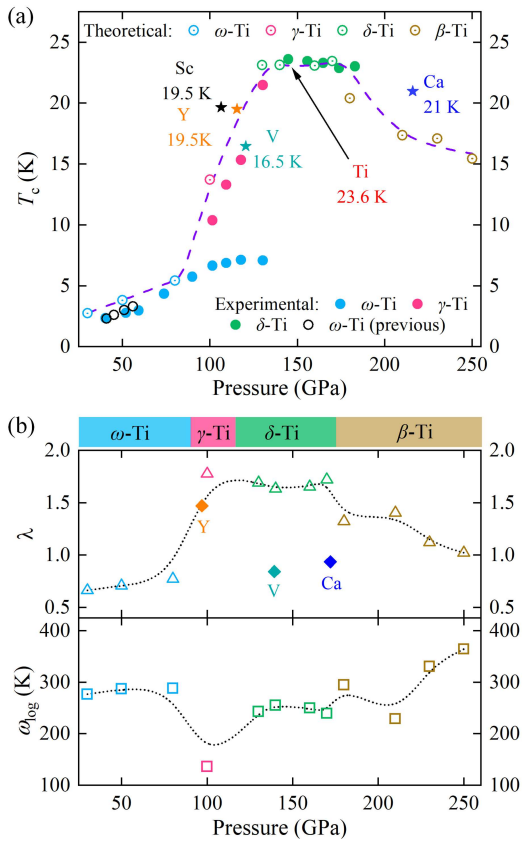}
            \caption{
                (a) DFT Calculation on $T_\mathrm{c}$ vs. pressure for Ti up to 250 GPa.
                The calculated $T_\mathrm{c}$ values are in good agreement with previous
                and present experimental results. Solid circles: $\omega$,
                $\gamma$, and $\delta$-Ti (experimental, this work), 
                Dot-filled circles: $\omega$, $\gamma$, and $\delta$-Ti (theoretical, this work), 
                and open circle: $\omega$-Ti (Ref.~\cite{BASHKIN200712}). The
                $T^\mathrm{max}_\mathrm{c}$  values for Sc, Y, V, and Ca are from
                Ref.~\cite{PhysRevB.78.064519,HAMLIN200782,PhysRevB.61.R3823,PhysRevB.83.220512}
                and more $T_\mathrm{c}$ ($P$) data can be found in Fig.~S2
                \cite{supp}. (b) Pressure dependence of EPC parameter
                $\lambda$ and logarithmic average frequency $\omega_\mathrm{log}$ calculated using the BCS
                and Migdal-Eliashberg theories framework
                \cite{bardeen1957theory,giustino2017electron}. The $\lambda$ values
                of Y, V, and Ca are taken from previous literature
                \cite{PhysRevLett.109.157004,Aftabuzzaman_2011,PhysRevB.67.094509}.
            }\label{fig2}
\end{figure}

The summarized $T_\mathrm{c}$ vs.~pressure in the pressure-temperature
($P$-$T$) phase diagram
is shown in Fig.~\ref{fig2}(a). Initially, the $T_\mathrm{c}$ exhibits a slow and nearly linear
increase from 2.3~K at $\sim$40~GPa to 7.1~K at $\sim$130~GPa. This demonstrates the
previous assumption that the $T_\mathrm{c}$ for Ti can increase linearly to
about 8.7~K when it transforms to the orthorhombic $\gamma$-phase at $\sim$128
GPa \cite{BASHKIN200712}. Until $\sim$145~GPa, the $T_\mathrm{c}$ rises rapidly to 23.6~K
with $dT_\mathrm{c}$/$dP$ = 0.29 K/GPa. The $T_\mathrm{c}$ value of 23.6~K is the highest among TMs (see
the $T^\mathrm{max}_\mathrm{c}$ of Sc, Y, V, and Ti in Fig.~\ref{fig2}(a)). Note that Ca shows a superconductivity
fluctuation at 29~K, but the rapid drop in resistance occurs at 21~K
\cite{PhysRevB.83.220512,PhysRevB.84.216501}.
Therefore, the $T^\mathrm{max}_\mathrm{c}$ = 23.6~K in Ti sets an exciting new record among elemental
superconductors. With further increasing pressure up to 183~GPa, the
$T_\mathrm{c}$ remains almost constant at 23~K. Such a robust $T_\mathrm{c}$ surviving over megabar
pressure is also observed in some Ti-bearing alloys, such as commercial NbTi
wire \cite{GuoAM2019} and high-entropy alloy (TaNb)$_{0.67}$(HfZrTi)$_{0.33}$ \cite{Guo13144}. 

The $T_\mathrm{c} (P)$ for Y, V, and Ca show monotonously increasing behavior without
interruption through the phase boundary (see Fig. S2) \cite{supp}. For Sc, the
appearance of the Sc-III phase causes the $T_\mathrm{c}$ to reduce significantly, but it
still maintains a positive $dT_\mathrm{c}$/$dP$ with further compression
\cite{PhysRevB.78.064519}. In contrast, the $T_\mathrm{c}$ of Ti is more sensitive to the changes in the crystal structure. The
pressure-dependent $T_\mathrm{c}$ matches well with the structural phase transition (SPT)
sequence, which characterizes a different $dT_\mathrm{c}$/$dP$ and upper critical field
$\mu_0H_\mathrm{c2}$ at 0~K (see Fig.~S3), estimated by the Werthamer-Helfand-Hohenberg
equation \cite{supp,PhysRev.147.295}. 

We further performed density functional theory (DFT) calculations to elucidate
the experimental observations in Ti. First, we theoretically confirmed the
previously reported SPT sequence
\cite{PhysRevLett.87.275503,PhysRevLett.86.3068,PhysRevB.67.132102,
PhysRevB.69.184102,HAO20101473,PhysRevB.91.134108,doi:10.1063/5.0014766}. 
The relative enthalpies vs. pressure
of the overall phases are shown in Fig.~S4 of the SM \cite{supp}. Our calculated
results indeed confirm a reported fact that Ti undergoes the 
$\alpha$ $\rightarrow$ $\omega$ $\rightarrow$ $\gamma$ $\rightarrow$ $\delta$
SPT sequence under pressure
\cite{PhysRevLett.87.275503,PhysRevLett.86.3068,PhysRevB.67.132102,
PhysRevB.69.184102,HAO20101473,PhysRevB.91.134108,doi:10.1063/5.0014766}. 
It is worth noting that the $\delta$ phase can relax to the $\beta$ phase at
pressures P $>$ 170 GPa, 
resulting in identical enthalpies for the $\delta$ phase and $\beta$ phase at 170-250 GPa. 

According to the McMillan-Allen-Dynes formula
\cite{PhysRev.167.331,PhysRevB.12.905}, the $T_\mathrm{c}$ value can be
estimated using three parameters: the effectively screened Coulomb repulsion
constant $\mu^{*}$, logarithmic average frequency $\omega_\mathrm{log}$, and EPC parameter 
$\lambda$. Here, $\mu^{*}$ is fixed at 0.19 (See Ref. \cite{supp}).
Figure~\ref{fig2}(a) plots the trend of the calculated $T_\mathrm{c}$,
which matches with the experimental results surprisingly well.
Particularly, the $\delta$ phase is predicted to host the most stable structure with
a $T_\mathrm{c}$ of 23~K between 130 and 170~GPa. However, after entering the
bcc $\beta$ phase, the $T_\mathrm{c}$ begins to decrease when P $>$ 180 GPa. Similar behavior is also observed
in superconducting Zr under pressure \cite{doi:10.1143/JPSJ.59.3843}.

The calculated $\omega_\mathrm{log}$ and $\lambda$ data are shown in
Fig.~\ref{fig2}(b). When $\gamma$-Ti appears at 100 GPa, $\omega_\mathrm{log}$ suddenly decreases. 
This abnormal frequency softening usually induces a sizeable enhancement of
$\lambda$ \cite{doi:10.1063/5.0033143}, leading to the increase of $T_\mathrm{c}$ in $\gamma$-Ti. 
The calculated $\lambda$ $\sim$ 1.65 for $\delta$-Ti demonstrates that it is a strongly
coupled superconductor at 130-170~GPa. The $\lambda$ value for $\delta$-Ti is the largest
among the surrounding elements near the pressure maximizing $T_\mathrm{c}$.
Experimentally, the $T_\mathrm{c}$ reaches a record value of about 23.6~K at
144.9~GPa, verifying that the high $T_\mathrm{c}$ of Ti is mainly contributed by strong EPC. Above
180 GPa, the $\omega_\mathrm{log}$ of the $\beta$ phase rises sharply with further pressure increase.
This drastic phonon hardening causes weakening of the electron-phonon
interactions, usually accompanying a decline in EPC parameter $\lambda$ (as shown in
Fig.~\ref{fig2}(b)). Recent work has claimed to observe the $\beta$-Ti phase
at 243~GPa \cite{doi:10.1063/5.0014766}. Given that the present pressure range
extends to 183~GPa, it remains uncertain
whether the robust superconductivity against larger volume shrinkage will
persist beyond 183 GPa. However, our theoretical prediction of the $\beta$-Ti phase
suggests a negative expectation. The robust superconductivity over megabar
pressure is proposed to be extremely unusual and virtually unique among known
superconductors \cite{GuoAM2019,Guo13144,PhysRevMaterials.3.090301}. 
Nevertheless, Jasiewicz et al. show that the EPC mechanism can
explain the experimental observations in (TaNb)$_{0.67}$(HfZrTi)$_{0.33}$
\cite{PhysRevB.100.184503}. Another theoretical calculation reveals that the robust superconductivity against the
large volume shrinkage is associated with the stability of the partial density
of states (DOS) contributed by the  and  orbital electrons from all
constituent atoms \cite{PhysRevMaterials.4.071801}, which remain almost unchanged in the
(TaNb)$_{0.67}$(HfZrTi)$_{0.33}$ and NbTi alloys. 

It is well known that both the EPC strength and the DOS around the Fermi level
($E_\mathrm{F}$) play important roles in the $T_\mathrm{c}$ enhancement in phonon-mediated
superconductors. We calculated pressure-dependent
DOS at Fermi level $N(E_\mathrm{F})$, as shown in Fig.~S5 \cite{supp}. 
The overall $N(E_\mathrm{F})$ decreases as pressure increases, but it reaches a local maximum 
at the $\gamma$ $\rightarrow$ $\delta$ phase transition. Combined with the increase in the EPC strength at the 
$\omega$ $\rightarrow$ $\gamma$ phase transition and a high plateau in
$\delta$ phase (see Fig.~\ref{fig2}(b)), the $T_\mathrm{c}$ gives the highest
value 23.6~K at the $\gamma$ $\rightarrow$ $\delta$ phase transition pressure and maintains
it around 23~K for the rest of $\delta$ phase. This powerfully demonstrates that the
high and robust $T_\mathrm{c}$ mainly arises from the enhanced EPC under pressure.

\begin{figure}[thp]
        \includegraphics[width=0.43 \textwidth]{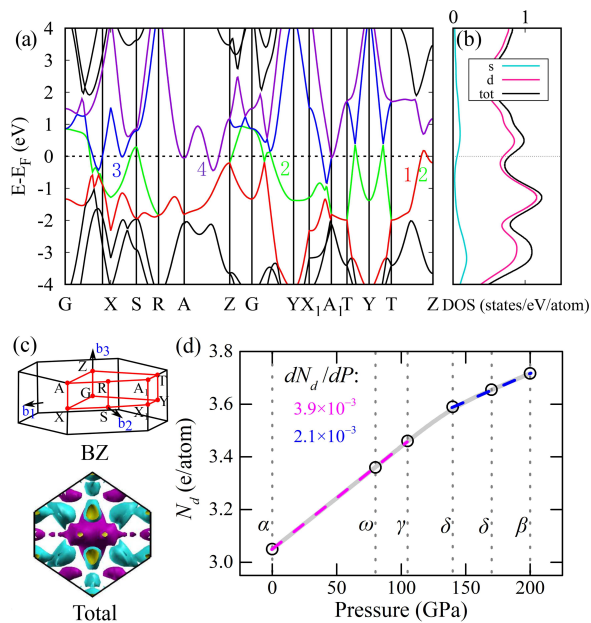}
            \caption{
                (a) Band structure and (b) Projected DOS of $\delta$-Ti at 140 GPa.
                (c) The first Brillouin zone and total Fermi surface. (d)
                Charge number of $d$ orbitals for Ti as a function of pressure.
                The results at 0, 80, 105, 140, and 200~GPa are obtained in
                the $\alpha$,$\omega$, $\gamma$, $\delta$ and $\beta$ phases, respectively.
            }\label{fig3}
\end{figure}

We took $\delta$-Ti at 140 GPa as a representative case and calculated its band
structure and orbital projected DOS to simulate the pressure-induced Tc
enhancement mechanism. As shown in Fig.~\ref{fig3}(a-b), four bands denoted by
No.~1$\sim$4 cross the $E_\mathrm{F}$ along the high-symmetry $k$ path in the Brillouin zone (BZ),
indicating metallic nature in the $\delta$-Ti phase. Note that band 1 and band 2
degenerate along the Z-T path. From the DOS and the orbital projected band
structure shown in Fig.~S6 of the SM \cite{supp}, the Ti-$d$ states mainly contribute
to the bands around the $E_\mathrm{F}$. Thus, the main physics of $\delta$-Ti is essentially
associated with the Ti-d orbitals, and $N(E_\mathrm{F})$ is about 0.85 states/eV per atom.
To establish the origin of the high $T_\mathrm{c}$ of $\delta$-Ti, we calculated its Fermi
surface (FS) at 140 GPa, as shown in Fig.~\ref{fig3}(c). The FS is composed of four
band sections, and their detailed descriptions are shown in Fig.~S7 of the SM
\cite{supp}. One key finding is that the Fermi-surface nesting appears in some Fermi
pockets, which substantially enhances the EPC and results in
high-$T_\mathrm{c}$ superconductivity in $\delta$-Ti
\cite{PhysRevLett.96.047004,2013Formation}.

Electron transferring from the s band to d band under pressure is well known
as a common feature of transition metals in many theoretical calculations
\cite{PhysRevB.78.064519,PhysRevB.67.094509,PhysRevB.31.1909}. 
Following a similar approach, our extended l$\ddot{\text{o}}$wdin charge analysis
also reveals a pressure-driven $s \rightarrow d$ transfer in Ti (see Fig. 3(d)). 
The increase in $N_{d}$ with pressure is caused by a relative increase in the energy of
the $s$-electrons compared to the $d$-electrons with pressure increase or volume
reduction \cite{PhysRevLett.86.3068}. We note that the pace of the $N_{d}$ increase slows down after
entering the $\gamma$ phase. By fitting with two linear regions before and
after 140~GPa, the change in $dN_{d}$/$dP$ is evident and almost halved. The equations of state
have a response to this dip: the total volume reduction from $\omega$ to
$\delta$ phase is 3.0\% at 147~GPa \cite{PhysRevLett.87.275503}. Besides, the record-high
$T_\mathrm{c}$ was experimentally observed at 144.9~GPa. Hence, this dip in
$dN_{d}$/$dP$ may approach the completion of the $s \rightarrow d$ transfer, and the record-high 
$T_\mathrm{c}$ is reached as we expect. Overall, by combining experiments with theoretical calculations, we demonstrate the
connection between the $s \rightarrow d$ transfer and superconductivity in Ti, calling for
electronic structure calculations to check whether the same scenario works in
other TMs with high $T_\mathrm{c}$.

\begin{figure}[thp]
        \includegraphics[width=0.43 \textwidth]{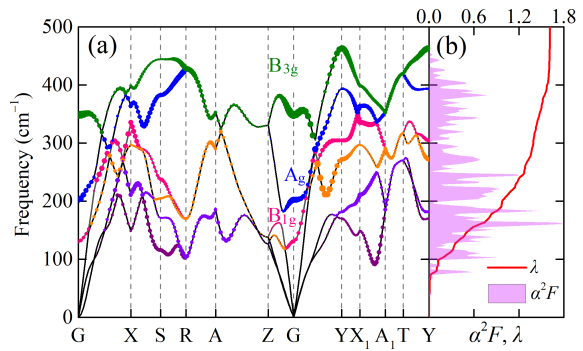}
            \caption{
                (a) Phonon dispersion of $\delta$-Ti at 140 GPa. The different colors
                denote the different phonon branches. The dots are
                proportional to the strengths of the phonon linewidth. (b)
                Eliashberg spectral function $\alpha^{2}F(\omega)$  and cumulative
                frequency-dependent EPC function $\lambda(\omega)$.
            }\label{fig4}
\end{figure}

The phonon spectrum, Eliashberg spectral function $\alpha^{2}F(\omega)$ and
cumulative $\lambda(\omega)$ of $\delta$-Ti were calculated to investigate the lattice dynamics and
electron-phonon interactions of the $\delta$ phase, as shown in
Fig.~\ref{fig4}. The absence of imaginary phonon modes in the whole BZ demonstrates its thermodynamical
stability at 140 GPa, which agrees with previous studies
\cite{PhysRevB.91.134108} . The phonon linewidth (denoted by the dots in Fig.~\ref{fig4}) for the phonon modes is plotted in
the phonon dispersion curve to gain further insights into the nature of the
EPC. The main considerable contribution to the EPC strength is the optical
branches based on the calculated phonon linewidths. In detail, we divide the
whole phonon frequencies into four intervals, namely,
0$-$120, 121$-$200, 201$-$300 and 301$-$500 cm$^{-1}$. 
The first interval corresponds to acoustic branches, and the
last three correspond to optical branches. Note that the optical branches are
dominated by three Raman modes designated by B$_\mathrm{1g}$, A$_\mathrm{g}$
and B$_\mathrm{3g}$. The results show that the low-frequency (below 120 cm$^{-1}$) vibration 
only contributes 20.2\% of the total EPC constant $\lambda$, which is largely caused by the softening of the
lowest acoustic branch along the $S$-$R$-$A$, $Z$-$G$, and $X_1$-$Y$ paths. This means that
the dominant contributions to $\lambda$ stem from the medium- and high-frequency
vibrations. It is consistent with the fact that the frequency of the highest
optical mode B$_\mathrm{3g}$ provides the largest linewidth around the $G$ point, followed
by the A$_\mathrm{g}$ and B$_\mathrm{1g}$ modes around the $G$ point. All Raman modes exhibit phonon
softening along other high-symmetry $k$ paths, indicating a strong EPC in
$\delta$-Ti.
Unlike the simple alkaline (earth) metals, in which $N(E_\mathrm{F})$ is
dominated by $s$ states, the $T_\mathrm{c}$ of
TMs usually exhibits a highly nonlinear dependent $T_\mathrm{c}$ on pressure
\cite{PhysRevB.78.064519}. Such complexity is associated with the nature of their partially filled d electrons
and SPT under pressure \cite{PhysRevB.78.064519,PhysRevB.98.214116}, which is consistent with our results.

In summary, we report the observation of a record high $T_\mathrm{c}$ of 23.6~K and robust
superconductivity in the $\delta$-Ti phase between 144.9 - 183~GPa. The unusual
superconductivity in pressurized Ti can be explained by the scenario of the
strong electron-phonon coupling effect from Fermi nesting formed by hole-like
and electron-like Fermi pockets and the substantial phonon softening of its
optical modes. Our results provide an in-depth insight into understanding the
pressure-tuning superconductivity of transition metals, which is fundamentally
important for the design and synthesis of high-$T_\mathrm{c}$ titanium alloy
superconductors for applications at extreme conditions.

The authors thank Dr. Zhipeng Yan, Dr. Cheng Ji, Dr. Junyue Wang, and Dr.
Dayong Liu for their technical support and theory guidance. We acknowledge
support from the National Natural Science Foundation of China Grant No.
U1930401, 12074153, and No.11674131. The authors appreciate Ms. Freyja O'Toole
for her language assistance.

%

\end{document}